\documentclass{article}
\usepackage{naturetex}
\usepackage{amsmath}
\usepackage{cite}
\usepackage{lineno}
\begin{document}
\maketitle
{\setstretch{1.5}
   
\section*{\centering Abstract}
Erasures are more favorable for quantum sensing than unflagged errors such as Pauli errors. However, realistic sensing noise does not usually appear as erasures; it often acts within the same sensing Hilbert space as the signal, making it difficult to identify and mitigate. For such noise, we establish a noise-model-agnostic necessary and sufficient condition for erasure conversion, identifying the noise components that can be converted into erasures and removed without damaging the signal. For components satisfying the condition, conversion can be realized by a passive dimension-lifted scheme requiring neither detailed noise knowledge nor active control. Theoretically, the protocol remains effective over a broad range of noise strengths and approaches the corresponding precision limit. Experimentally, in single-photon phase sensing, we recover standard-quantum-limit precision in a Pauli-noise channel with erasure-convertible weight 0.5, using orbital angular momentum as the ancilla. These results provide a practical route to robust quantum sensing under realistic noise.
}
\titleformat{\section}[block]  
  {\normalfont\bfseries\Large}  
  {\thesection}{1em}{}
\section*{Introduction}
Quantum sensing enables precision beyond classical strategies by exploiting quantum coherence and entanglement~\cite{Giovannetti2006,Pezze2018}.
However, this quantum advantage is fundamentally challenged by unavoidable environmental noise, which induces decoherence and limits the achievable precision~\cite{Escher2011NatPhys,DemkowiczDobrzanski2012NatCommun}.

A variety of strategies have been developed to protect quantum sensing from noise, including quantum error correction~\cite{Kessler2014,Dur2014,Unden2016PRL,Wang2022RadiometryQEC,Zhou2020AQEC}, dynamical decoupling~\cite{Viola1999,Sekatski2016DDScaling,Merkel2021DDEnsembles,Louzon2025CPDD}, and adaptive estimation schemes~\cite{Hentschel2011EfficientAdaptive,DemkowiczDobrzanski2017AdaptiveMarkovian,Kurdzialek2023AdaptivenessCausal,Pang2017OptimalAdaptiveControl}. 
These active approaches have achieved remarkable progress, but they often require prior noise-model information, accurate controls, or additional ancilla qubits, thereby imposing a costly overhead. Moreover, their performance is limited when the noise channel is unknown, imperfectly characterized, drifts over time, or when the control and ancilla are imperfect~\cite{Louzon2025CPDD,Degen2017QuantumSensing,Shettell2021PracticalQEC}.

A complementary approach to noise tolerance, widely explored in quantum computing, is erasure conversion, namely the process of converting errors into erasures~\cite{Wu2022ErasureConversion,Scholl2023ErasureConversion,Ma2023MidCircuitErasure}. Erasures are errors whose occurrence and location can be detected~\cite{Bennett1997Erasure,Grassl1997Erasure}. Compared with unflagged errors such as Pauli errors, erasures are easier to handle and can increase error-correction thresholds or reduce overhead~\cite{Kang2023IonErasureQEC}. Erasures are also favorable for quantum sensing: when the dominant noise is erasure-type noise, the precision bound can be improved~\cite{Eraqubit2024}. However, most sensing noise does not naturally appear in erasure form, but instead often coexists with the signal within the same sensing Hilbert space. Unlike loss or leakage errors, which drive the state out of the sensing subspace and can therefore be more readily identified as, or converted into, erasures, such noise remains hidden within the sensing subspace, making it harder to identify and suppress. We refer to this class of noise as general in-space noise.

Converting general in-space sensing noise into erasures is challenging and has not been systematically addressed, since the noise must be identified and removed while preserving the signal. This leads to a fundamental question: what is the boundary of erasure conversion for general in-space noise in quantum sensing? Specifically, which noise components can be converted into erasures, and which components cannot be converted without damaging the signal? In addition, for practical sensing applications, it is desirable to realize such conversion passively, thereby reducing the resource overhead.

In this paper, we establish a necessary and sufficient erasure-conversion condition for general in-space quantum sensing noise under passive erasure conversion and experimentally demonstrate the conversion. The condition determines an erasure-convertible sector and a non-convertible sector of a general in-space noise channel. It is determined by the sensing generator and is independent of the specific noise model. Guided by this condition, we construct a static dimension-lifted scheme that converts all erasure-convertible components into erasures. The converted erasures are then removed by a final ancilla projection, since in quantum sensing it is often sufficient to identify and discard erroneous contributions so that they do not contaminate the data used for parameter estimation. The scheme is noise-model-agnostic in execution, requires neither active control nor precise timing, can use intrinsic degrees of freedom of the probe as the ancilla, and remains tolerant to a certain level of ancilla noise. These features make it more robust against temporal noise fluctuations, and less resource-demanding.
We derive the effective quantum Fisher information (QFI) of the framework and validate its performance through representative examples. The results show that our framework can recover lost precision over a broad range of noise strengths, bringing the achievable precision close to the corresponding theoretical precision limit. Experimentally, we demonstrate single-photon phase sensing in a Pauli-noise channel with total erasure-convertible weight 0.5, using orbital angular momentum (OAM) as the ancilla. The measured precision is consistent with the theoretical prediction and is significantly closer to the standard-quantum-limit (SQL) compared with the untreated noisy case. These results show that our work provides a practical route to robust quantum sensors, with potential applications in biomedicine, environmental monitoring, and related fields.

\section*{Results}
\subsection*{In-space noise and passive erasure conversion scheme for quantum sensing}
Consider a generic sensing scenario where an unknown parameter $\theta$ is encoded in an initial probe state $\rho_0$ in a $d$-dimensional sensing space $\mathcal H_S$. The parameter is encoded through the unitary evolution $\Lambda_\theta(\rho)=U_\theta \rho U_\theta^\dagger$, with $U_\theta=e^{-i\theta G}$, where $G$ is the sensing generator. In the presence of noise, the evolution is described by the noisy sensing channel $\mathcal E_\theta=\mathcal N\circ\Lambda_\theta$, where $\mathcal N$ is the noise channel acting during the sensing process.

\begin{figure}[H]
    \centering
    \includegraphics[width=1\linewidth]{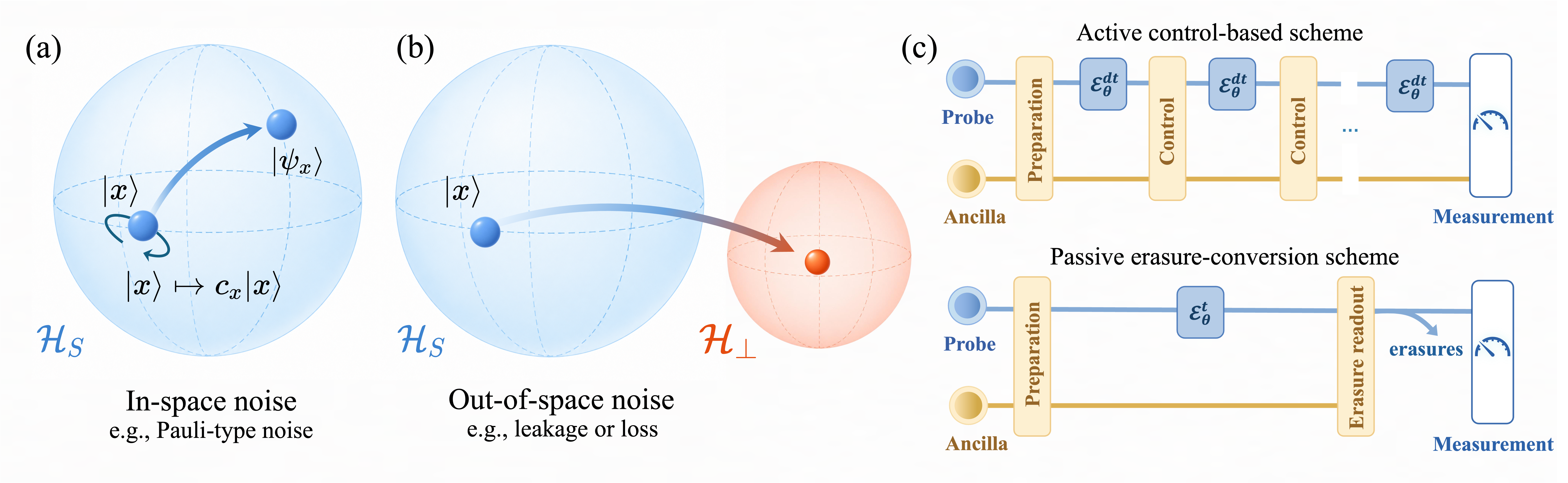}
    \caption{\textbf{Comparison of noise types and sensing schemes.}
    \textbf{(a)} In-space noise acts within the sensing Hilbert space $\mathcal H_S$. It may change the phase or amplitude of a state, or map it to another state within $\mathcal H_S$, making the noise contribution hidden in the same space as the signal.
    \textbf{(b)} Out-of-space noise, such as leakage or loss, maps the state outside the sensing Hilbert space, and therefore naturally carries a detectable signature, making it easier to identify as, or convert into, erasures.
    \textbf{(c)}
    Comparison between an active control-based scheme and the passive erasure-conversion scheme considered here. Active schemes rely on repeated, fast, and precise control operations during the sensing evolution. In contrast, the passive erasure-conversion scheme requires only static operations. An ancilla is introduced to encode the probe, and the erasure readout removes the converted erasures from the final retained output.
    }
    \label{fig:concept}
\end{figure}

The noise can be classified according to whether it takes the state out of the sensing space. Leakage, loss, and other out-of-space noise map the state from $\mathcal H_S$ to an external orthogonal space $\mathcal H_\perp$. The resulting states lie outside the original sensing space and often carry an external signature that can be detected. Therefore, such noise events can be more readily identified as, or converted into, erasures, as illustrated in Fig.~\ref{fig:concept}(b). By contrast, the general in-space noise considered in this work acts entirely within the sensing Hilbert space, taking the form $\mathcal N:\mathcal L(\mathcal H_S)\rightarrow\mathcal L(\mathcal H_S)$. It may modify the phase or amplitude of the state, as represented by $|x\rangle \mapsto c_x |x\rangle$ with $c_x\in \mathbb{C}$, or it may map the state to another state within $\mathcal H_S$, as represented by $|x\rangle \mapsto |\psi_x\rangle \in \mathcal H_S$ with $|\psi_x\rangle=\sum_i c_i^{(x)}|x_i\rangle$.
Unlike out-of-space noise, in-space noise remains hidden in the same Hilbert space as the signal and does not carry a native erasure flag. This makes it difficult to directly identify and convert into erasures (Fig.~\ref{fig:concept}(a)).

Here we aim to determine how in-space noise can be converted into erasures and what the boundary of such conversion is. In contrast to most noise-resilient schemes, such as quantum error correction, which are active and require substantial overhead, we focus on a passive approach (Fig.~\ref{fig:concept}(c)). In this passive erasure-conversion scheme, no active operations are applied to process the noise during sensing. An ancilla is introduced to encode the probe. This ancilla can exploit intrinsic degrees of freedom of the probe, reducing resource cost and tolerating a certain level of noise. The erasure readout removes the converted erasures from the final retained output. These points will be discussed in detail below.

\subsection*{Erasure-conversion condition}

Addressing general in-space noise requires a unified framework for noise modeling and treatment, which enables passive erasure conversion to execute in a noise-model-agnostic way.
Most noise-resilient studies have been based on Kraus or Lindblad operator representations. These representations are useful for specific noise models, but they are not unique and therefore less suitable for a general framework for arbitrary in-space noise~\cite{NielsenChuang2010,Unden2016PRL,Zhou2018NatComm}.
Instead of directly analyzing the noise operators, we re-express a general in-space noise channel $\mathcal{N}(\cdot)=\sum_r K_r (\cdot) K_r^\dagger$ with respect to $G$, and on this basis give the necessary and sufficient condition specifying which in-space noise components can be converted into erasures and which cannot under the passive scheme. This condition is called the erasure-conversion condition in this paper.

Let $\{\ket{x}\}$ be the eigenbasis of $G$. This eigenbasis defines the sensing basis, and $x$ labels the corresponding basis states. The operator space $\mathcal L(\mathcal H_S)$ then decomposes into two Hilbert--Schmidt orthogonal subspaces,
\begin{equation}
\mathcal D_G:=\mathrm{span}\{\ket{x}\bra{x}\}_{x},\quad
\mathcal O_G:=\mathrm{span}\{\ket{x}\bra{y}\}_{x\neq y},\quad
\mathcal L(\mathcal H_S)=\mathcal D_G\oplus\mathcal O_G .
\end{equation}

\begin{figure}[H]
  \centering
  \includegraphics[width=0.9\linewidth]{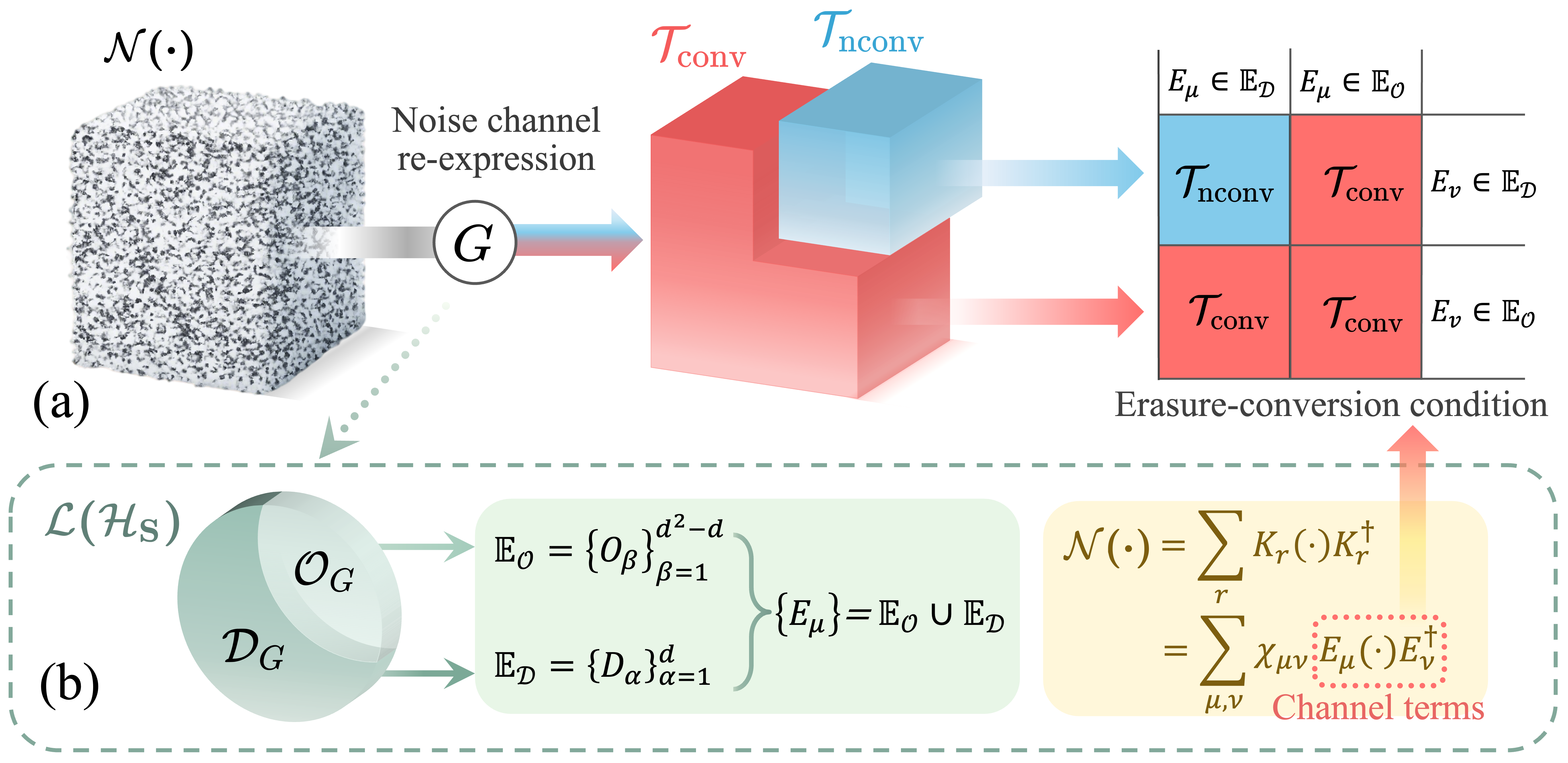}
    \caption{\textbf{Schematic of the general in-space noise channel re-expression and erasure-conversion condition.} 
\textbf{(b)} A general in-space noise channel $\mathcal{N}(\cdot)$ is re-expressed with respect to the sensing generator $G$. In the fixed sensing basis defined by $G$, the operator space $\mathcal{L}(\mathcal{H}_S)$ is decomposed into the diagonal subspace $\mathcal{D}_G$ and the off-diagonal subspace $\mathcal{O}_G$, providing the Hilbert\text{-}Schmidt orthonormal operator bases $\mathbb{E}_{\mathcal D}$ and $\mathbb{E}_{\mathcal O}$. In the operator basis $\{E_\mu\}$, a general in-space noise channel is re-expressed as $\mathcal N(\cdot)=\sum_{\mu,\nu}\chi_{\mu\nu}E_\mu(\cdot)E_\nu^\dagger$. The analysis therefore focuses on the channel terms $E_\mu(\cdot)E_\nu^\dagger$ (highlighted by the dashed red box), rather than the noise operators. 
\textbf{(a)} Based on this re-expression, the erasure-conversion condition defines two distinct classes of in-space noise channel terms: the erasure-convertible set $\mathcal{T}_{\mathrm{conv}}$ (red) and the non-convertible set $\mathcal{T}_\mathrm{nconv}$ (blue). Specifically, a term is non-convertible if and only if both $E_\mu$ and $E_\nu$ belong to $\mathbb{E}_{\mathcal D}$; otherwise, if either operator lies in $\mathbb{E}_{\mathcal O}$, the term is convertible.}
    \label{fig:decomposition_principle}
\end{figure}

\noindent Here $\mathcal D_G$ contains operators that preserve the labels of eigenstates, whereas $\mathcal O_G$ contains operators that change them. 

We then choose complete Hilbert\text{-}Schmidt orthonormal bases for \(\mathcal D_G\) and \(\mathcal O_G\), denoted by
$\mathbb E_{\mathcal D}=\{D_\alpha\}_{\alpha=1}^{d},  \mathbb E_{\mathcal O}=\{O_\beta\}_{\beta=1}^{d^2-d}$.
Accordingly,
$\{E_\mu\}=\mathbb E_{\mathcal D}\cup\mathbb E_{\mathcal O}$ forms a complete operator basis for $\mathcal L(\mathcal H_S)$.
In this basis, a general in-space noise channel can be written as
$\mathcal N(\cdot)=\sum_{\mu,\nu}\chi_{\mu\nu}E_\mu (\cdot) E_\nu^\dagger$ (see Methods).
The analysis therefore focuses on the channel terms $E_\mu(\cdot)E_\nu^\dagger$, rather than on a particular set of noise operators, as schematically illustrated in Fig.~\ref{fig:decomposition_principle}.

\textbf{Erasure-conversion condition.} Under passive erasure conversion, a channel term $E_\mu(\cdot)E_\nu^\dagger$ is erasure-convertible if and only if at least one of its left and right operators belongs to $\mathcal O_G$. 
Equivalently, the erasure-convertible and non-convertible sets are
\begin{align}
\mathcal T_\mathrm{conv}
&:=\big\{\,E_\mu(\cdot)E_\nu^\dagger\,\big|\,E_\mu\in\mathbb E_{\mathcal O}\ \text{or}\ E_\nu\in\mathbb E_{\mathcal O}\,\big\},\\
\mathcal T_\mathrm{nconv}
&:=\big\{\,E_\mu(\cdot)E_\nu^\dagger\,\big|\,E_\mu\in\mathbb E_{\mathcal D}\ \text{and}\ E_\nu\in\mathbb E_{\mathcal D}\,\big\}.
\end{align}
Thus, as shown in Fig.~\ref{fig:decomposition_principle}, a term is non-convertible precisely when both sides preserve the sensing-basis labels. If either side changes the label, the term is erasure-convertible. Within $\mathcal T_{\rm conv}$, terms of the form $E_\mu(\cdot)E_\mu^\dagger$ can be converted into detectable erasure events, whereas cross terms $E_\mu(\cdot)E_\nu^\dagger$ with $E_\mu\neq E_\nu$ represent coherent contributions and do not define independent outcome events. Nevertheless, these cross terms are removed by the same final readout and therefore do not affect parameter estimation. In this sense, they are also classified as erasure-convertible terms in this work. By contrast, the signal-preserving identity term \(I(\cdot)I\) is non-convertible and is therefore retained.

Crucially, the condition is dictated by the subspace decomposition \(\mathcal D_G\oplus\mathcal O_G\) induced by \(G\), rather than by any particular choice of the operator basis \(\{E_\mu\}\). Thus, we obtain the key theorem for erasure conversion in quantum sensing.

\textbf{Theorem I.} Consider a finite-dimensional sensing space $\mathcal H_S$ with sensing generator $G$, subject to an in-space noise channel $\mathcal N$ that keeps the probe within $\mathcal H_S$. Under a passive scheme, a noise-channel component can be removed from the retained sensing outcome as an erasure without damaging the signal if and only if it satisfies the erasure-conversion condition.

Furthermore, because the erasure-conversion condition depends on \(G\), in sensing tasks where physically equivalent working bases are available, one may in principle optimize the orientation of the generator so that more noisy terms fall into \(\mathcal T_\mathrm{conv}\), thereby further enhancing the sensing performance (see Methods).

\textbf{Qubit examples}
To illustrate Theorem~I, we consider a qubit probe with $G \propto \sigma_z$, so that the sensing basis is $\{\ket{0},\ket{1}\}$. The operator space decomposes into $\mathcal D_G = \mathrm{span}\{\ket{0}\bra{0}, \ket{1}\bra{1}\}$ and $\mathcal O_G = \mathrm{span}\{\ket{0}\bra{1}, \ket{1}\bra{0}\}$. Accordingly, we may choose $\mathbb E_{\mathcal D} = \{I, \sigma_z\}$ and $\mathbb E_{\mathcal O} = \{\sigma_x, \sigma_y\}$.

A general qubit in-space noise channel can be represented by Kraus operators $\{K_r\}$, each of which can be expanded in the Pauli basis as 
$K_r = n_{r,0} I + n_{r,x} \sigma_x + n_{r,y} \sigma_y + n_{r,z} \sigma_z$, where the coefficients are complex. Regardless of how many independent $K_r$ exist, if $n_{r,z} = 0$, the noise is fully erasure-convertible. For example, a single operator $K_1 = n_x \sigma_x + n_y \sigma_y$ or two operators $K_1 = \sigma_x$, $K_2 = \sigma_y$ are fully convertible. If $n_{r,z} \neq 0$, there are non-convertible residual noise terms involving $\sigma_z$.

For a single-qubit probe, the output after the noise channel consists of a combination of erasure-convertible and non-convertible parts. The coherence between these parts depends on the degree of decoherence of the noise channel. Upon performing a final joint measurement of the probe and ancilla, the state is projected into orthogonal outcomes. The retained outcomes, corresponding signal-preserving contribution \(I(\cdot)I\) and the non-convertible noise, are used to estimate the parameter $\theta$, while the erasure outcomes are discarded. 

\subsection*{Non-convertible components of in-space noise}

We prove the necessary part of Theorem~I in this section. Consider an in-space noise channel term \(E_\mu(\cdot)E_\nu^\dagger\), where \(E_\mu,E_\nu\in\mathcal D_G\). They can be written as \(E_\mu=\sum_x d_{\mu x}\ket{x}\bra{x}\) and \(E_\nu=\sum_x d_{\nu x}\ket{x}\bra{x}\). Acting on a density-matrix element \(\ket{x}\bra{x'}\), this term gives
$E_\mu\ket{x}\bra{x'}E_\nu^\dagger
=
d_{\mu x}d_{\nu x'}^* \ket{x}\bra{x'}$.
Therefore, the term only modifies the complex coefficient of the matrix element.

This structure is closely related to the signal evolution itself. The parameter is encoded by \(U_\theta\), which acts on \(\ket{x}\bra{x'}\) as
$\ket{x}\bra{x'}
\rightarrow
e^{-i\theta(g_x-g_{x'})}\ket{x}\bra{x'}$.
Thus, such a noise component produces the same type of output state as the signal evolution. It does not create any additional structure from which an erasure flag can be constructed. Under the passive scheme considered here, without active recovery, environment monitoring, feedback, or intermediate measurements, there is no extra information that can distinguish this noise contribution from the signal evolution at the final readout. Removing this class as erasures would also remove the signal-preserving contribution \(I(\cdot)I\). Therefore, terms \(E_\mu(\cdot)E_\nu^\dagger\) with \(E_\mu,E_\nu\in\mathcal D_G\) cannot be passively converted into erasure outcomes and must remain in the retained part.

\subsection*{Converting \(\mathcal T_\mathrm{conv}\) into erasures}
To prove the sufficient condition of Theorem~I, we explicitly construct a passive scheme that converts the terms in \(\mathcal T_\mathrm{conv}\) into erasures. Here, by erasures we mean that, after the final erasure readout, the converted channel terms have no support on the retained sensing outcome and are therefore removed from the data used for parameter estimation.

\begin{figure}[H]
  \centering
  \includegraphics[width=1\textwidth]{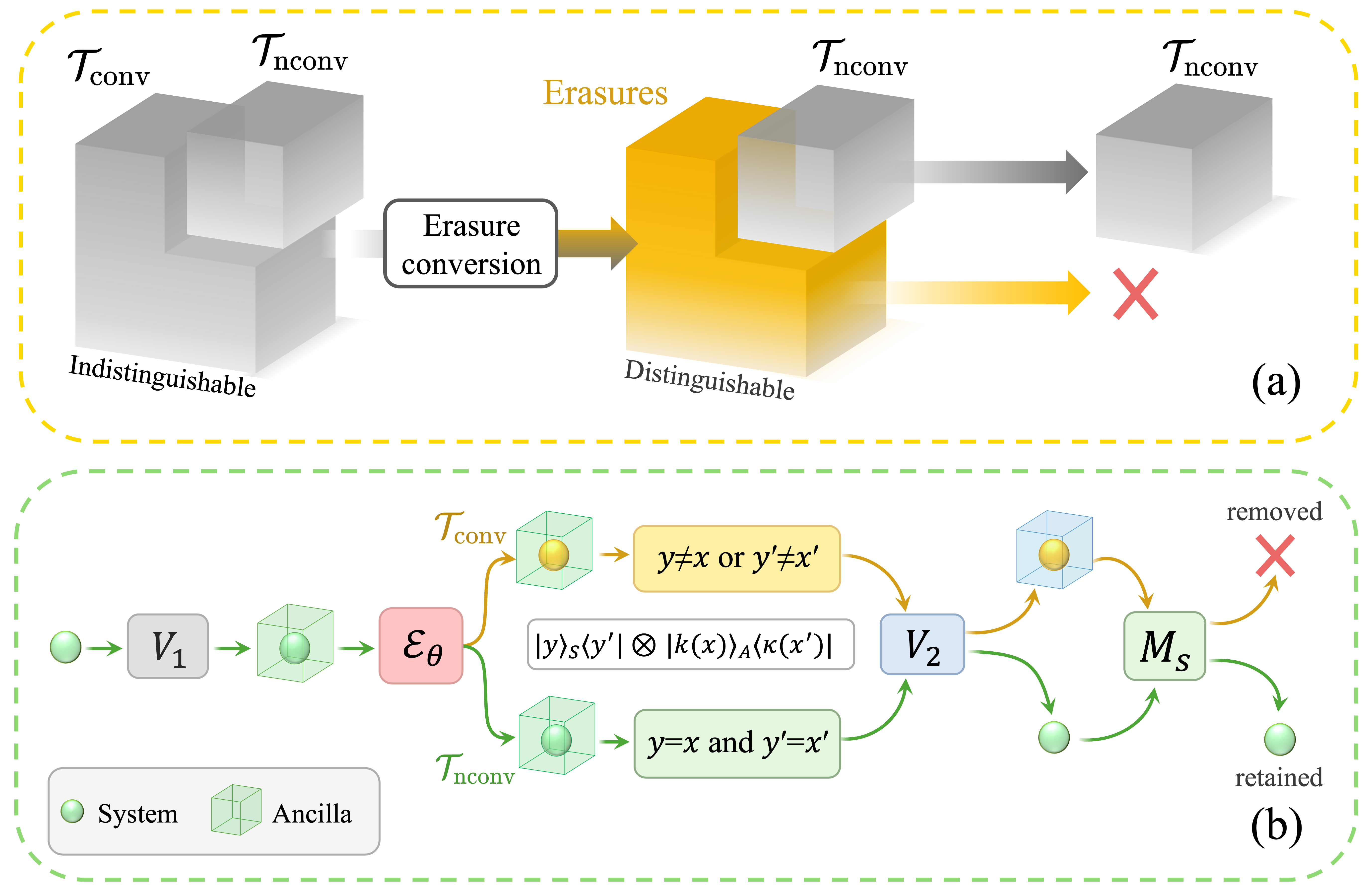}
\caption{\textbf{(a) Conceptual illustration and (b) workflow of the passive erasure conversion scheme.}
(a) Before erasure conversion, the convertible and non-convertible components of in-space noise are indistinguishable within the sensing subspace. After erasure conversion, the components in $\mathcal T_\mathrm{conv}$ are converted into erasures (highlighted in yellow) and removed from the retained sensing outcome by the final readout, leaving only $\mathcal T_\mathrm{nconv}$ in the retained outcome.
(b) The probe and ancilla are prepared into a joint state through $V_1$, where the sphere denotes the sensing system and the cube denotes the ancilla. The encoding $V_1$ assigns mutually orthogonal ancilla states to the sensing-basis states, thereby recording their basis indices in the ancilla. After the noisy evolution $\mathcal E_\theta$, the joint terms are transformed from
$|x\rangle_S\langle x'| \otimes |\kappa(x)\rangle_A\langle\kappa(x')|$
to
$|y\rangle_S\langle y'| \otimes |\kappa(x)\rangle_A\langle\kappa(x')|$. Here, $(x,x')$ denote the sensing-basis indices recorded in the ancilla during encoding, while $(y,y')$ denote the corresponding indices after the noisy evolution.
Terms in $\mathcal T_\mathrm{conv}$ induce sensing-basis changes that break the system\text{-}ancilla index correspondence, illustrated by the sphere changing from green to yellow. After decoding by $V_2$, mismatched terms have no support on the retained ancilla state $\ket{0}_A$ and are rejected by the final projection $M_s$.
}
\label{fig:architectures}
\end{figure}

The central idea is as follows. First, each sensing basis state is appended with an ancilla flag, and these ancilla flags are mutually orthogonal. After the action of noise, a fixed decoding operation checks whether the index of the system part still matches the index recorded in the ancilla. If a channel term belongs to \(\mathcal T_\mathrm{conv}\), it necessarily induces a transition between sensing basis states on at least one side of the density-matrix term, thereby breaking the initial correspondence. After decoding, such a mismatch is mapped outside the retained ancilla state and can therefore be removed uniformly by a single final ancilla projection, as illustrated in Fig.~\ref{fig:architectures}(a,b).
For clarity, the derivation below is presented for a noiseless ancilla, since our main target is the system noise. However, the scheme can also tolerate ancilla noise that likewise breaks the system\text{-}ancilla correspondence (see Supplementary).

Specifically, let \(\{|x\rangle_S\}_{x\in\mathcal X}\) be the sensing basis of \(\mathcal H_S\), where \(\mathcal X\) denotes the set of sensing-basis labels. We introduce an ancilla \(A\) indexed by \(\mathcal A\), and define an injective map \(\kappa:\mathcal X\to\mathcal A\), such that each \(|x\rangle_S\) is associated with an  ancilla state \(|\kappa(x)\rangle_A\), and the states \(\{|\kappa(x)\rangle_A\}_{x\in\mathcal X}\) are mutually orthogonal. For each \(\kappa(x)\), we define a unitary \(U_{\kappa(x)}\) acting on the ancilla such that $U_{\kappa(x)}|0\rangle_A=|\kappa(x)\rangle_A,
U_{\kappa(x)}^\dagger|\kappa(x)\rangle_A=|0\rangle_A$.
Then the encoding and decoding operations are defined, respectively, by
\begin{equation}
    V_1=\sum_{x\in\mathcal X}|x\rangle_S\langle x|\otimes U_{\kappa(x)},\quad
V_2=\sum_{x\in\mathcal X}|x\rangle_S\langle x|\otimes U_{\kappa(x)}^\dagger,
\end{equation}
which satisfy the identity $V_2V_1=\mathbb I_S\otimes \mathbb I_A$.
For the joint initial state $\rho_0\otimes|0\rangle_A\langle0|
=\sum_{x,x'}\rho_{xx'}|x\rangle_S\langle x'|\otimes|0\rangle_A\langle0|$, the encoded state becomes
$\rho_{\mathrm{enc}}
=\sum_{x,x'}\rho_{xx'}\,|x\rangle_S\langle x'|\otimes|\kappa(x)\rangle_A\langle\kappa(x')|$.
That is, the encoding operation writes the sensing-basis indices \((x,x')\) associated with each density matrix term into the ancilla.

Assuming that the ancilla is noiseless, the terms in the joint density matrix after $\mathcal E_\theta$ take the form

\noindent $|y\rangle_S\langle y'|\otimes|\kappa(x)\rangle_A\langle\kappa(x')|$,
where $(x,x')$ are the sensing-basis indices written into the ancilla during encoding, whereas $(y,y')$ are the corresponding indices of the system operator after $\mathcal E_\theta$.
After applying \(V_2\), the ancilla returns to \(|0\rangle_A\langle0|\) if and only if $y=x$ and $y'=x'$.
This follows from $\langle0|U_{\kappa(y)}^\dagger|\kappa(x)\rangle_A=\delta_{yx}$ and
$\langle\kappa(x')|U_{\kappa(y')}|0\rangle_A=\delta_{x'y'}$.

For any noise channel term in $\mathcal T_\mathrm{conv}$, the sensing-basis label changes on at least one side of the density-matrix term, thereby breaking the index correspondence established during encoding. We define the retained subspace after decoding as $\mathcal H_{\rm ret}=\mathcal H_S\otimes \mathrm{span}\{\ket{0}_A\}$, with the corresponding projector $M_s=\mathbb I_S\otimes\ket{0}_A\bra{0}$. The orthogonal subspace of this retained subspace is defined as the erasure subspace, $\mathcal H_{\rm er}=\mathcal H_S\otimes \left(\mathrm{span}\{\ket{0}_A\}\right)^\perp$. After decoding, the joint channel term corresponding to such a mismatch has no support on the retained part selected by $M_s$, that is,
\begin{equation}
M_sV_2\left[(E_\mu\otimes\mathbb I_A)\rho_{\mathrm{enc}}(E_\nu^\dagger\otimes\mathbb I_A)\right]V_2^\dagger M_s=0,
\quad
\text{if }E_\mu\in\mathbb E_{\mathcal O}\ \text{or}\ E_\nu\in\mathbb E_{\mathcal O}.
\end{equation}
Thus, every channel term in $\mathcal T_{\mathrm{conv}}$ is removed from the retained outcome. Among these terms,  converted $E_\mu(\cdot)E_\mu^\dagger$ terms correspond to erasure events in the erasure subspace. Converted cross terms $E_\mu(\cdot)E_\nu^\dagger$ with $E_\mu\neq E_\nu$ do not define independent outcome events, but they are also removed by the same final readout. Thus, they are also regarded as erasures in this work. By contrast, the identity term $I(\cdot)I$ is non-convertible and is preserved. Therefore, the scheme removes the convertible terms without discarding the signal.

This proves the sufficient direction of the erasure-conversion condition: all channel terms in $\mathcal T_{\mathrm {conv}}$ can be passively converted into erasures without damaging the signal. Moreover, this construction is $\theta$-independent and noise-model-agnostic in execution; it requires no intermediate recovery, adaptive control, or real-time feedback during sensing, and therefore simplifies implementation and reduces resource costs.

\subsection*{Theoretical QFI analysis}
Under the passive erasure-conversion framework, the attainable information is quantified by the effective QFI
\begin{equation}
\label{eq:main-FQ}
Q^{\mathrm{eff}}(\theta)
=
p_s(\theta)\,Q\!\big[\rho^{(s)}_\theta\big]
+
\frac{\big(\partial_\theta p_s(\theta)\big)^2}{p_s(\theta)\,[1-p_s(\theta)]},
\end{equation}
where $p_s(\theta)$ is the probability of the retained outcome and $\rho_\theta^{(s)}$ is the corresponding normalized retained state. When $p_s$ is independent of $\theta$, the classical Fisher term vanishes and the effective QFI becomes $Q^{\mathrm{eff}}(\theta)=p_s\,Q[\rho_\theta^{(s)}]$. If no non-convertible noise component remains in the retained outcome, the framework saturates the upper bound
\begin{equation}
\label{eq:FQ-upper}
Q^{\mathrm{eff}}(\theta)
\le
p_s(\theta)\,Q\!\big[\Lambda_\theta(\rho_0)\big]
+
\frac{\big(\partial_\theta p_s(\theta)\big)^2}{p_s(\theta)\,[1-p_s(\theta)]},
\end{equation}
which reduces to $Q^{\mathrm{eff}}(\theta)\le p_s\,Q[\Lambda_\theta(\rho_0)]$ when $p_s$ is $\theta$-independent. The detailed derivation is given in the Supplementary Material.

To illustrate these results for both Pauli and non-Pauli noise, we consider phase damping and amplitude damping as two representative examples. In the following examples, we take the sensing generator to be \(G=\sigma_z/2\) and use \(\ket{+}^{\otimes N}\) as the probe state.

\paragraph{Phase damping.}
We consider the phase-damping channel
$\mathcal{E}_p(\rho)=(1-p)\rho+p\,\sigma_z\rho\sigma_z$, with $p\in[0,\tfrac12]$.
When an equivalent sensing basis is available, the sensing generator can be chosen along an equivalent direction, such as $G=\sigma_y/2$, so that all noisy terms fall into $\mathcal{T}_{\mathrm{conv}}$.
Our framework then reaches the upper bound,
$Q_p^{\rm er}(N,p)=Q_p^{\rm ub}(N,p)=N(1-p)$.
The untreated noisy baseline is
$Q_p^{\rm noisy}(N,p)=N(1-2p)^2$.
\begin{figure}[H]
  \centering
  \newcommand{\subw}{0.48\linewidth}
  \begin{subfigure}[t]{\subw}
    \centering
    \includegraphics[width=\linewidth]{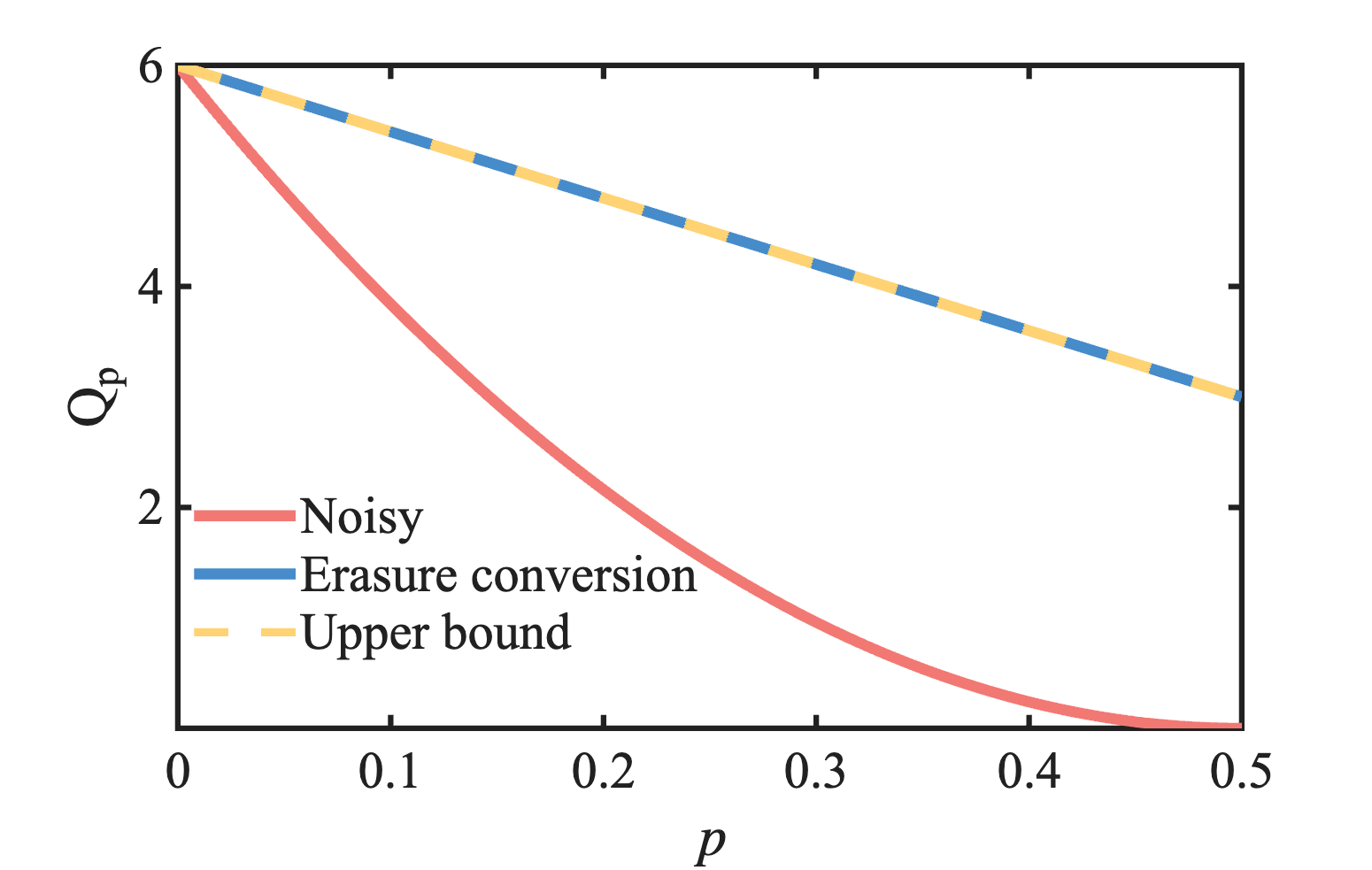}
    \caption{$Q_p$ versus $p$ at $N=6$.}
  \end{subfigure}\hfill
  \begin{subfigure}[t]{\subw}
    \centering
    \includegraphics[width=\linewidth]{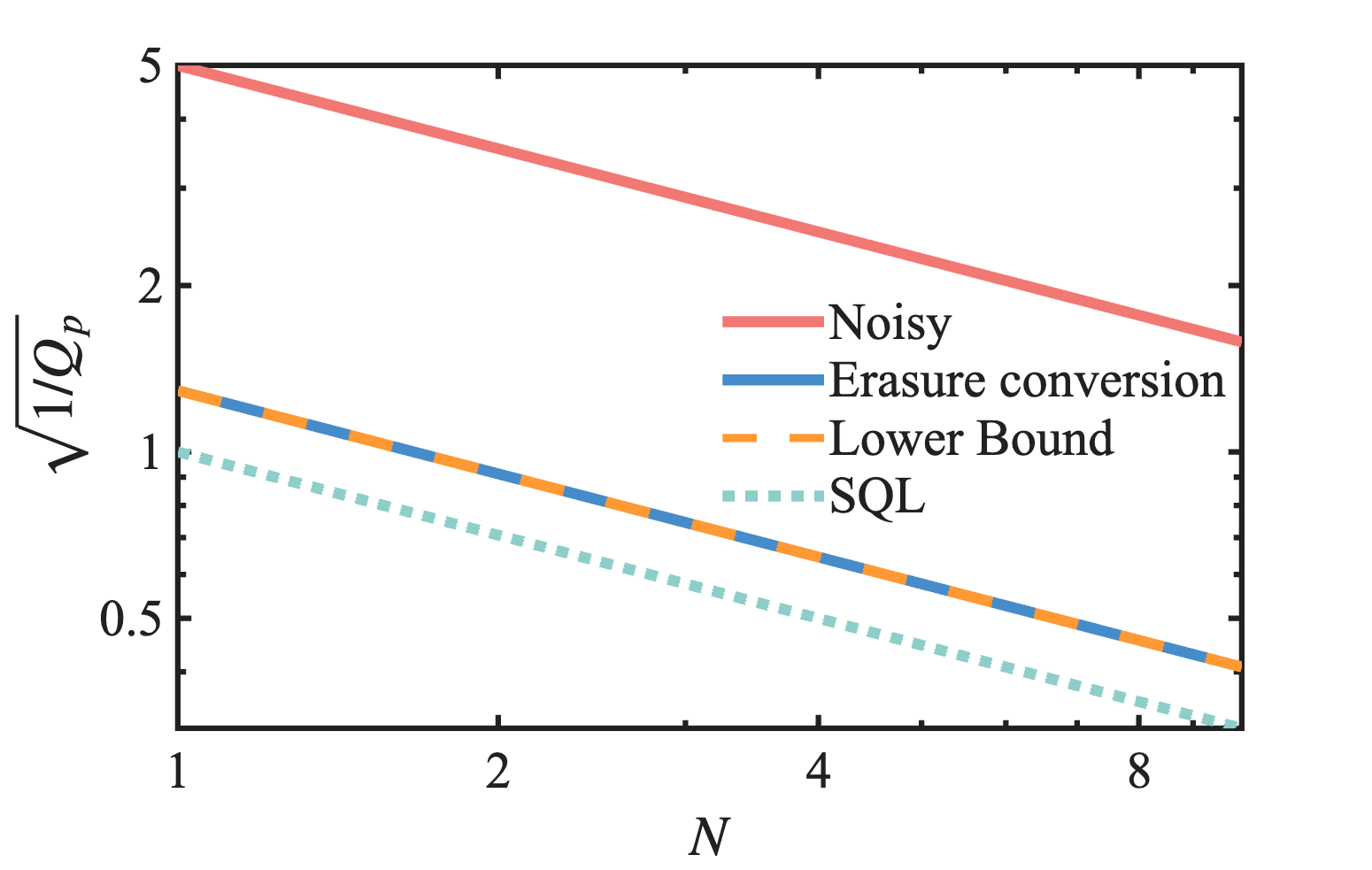}
    \caption{$\sqrt{1/Q_p}$ versus $N$ at $p=0.4$ on log\text{-}log axes.}
  \end{subfigure}
\caption{\textbf{Phase damping.}
The noisy baseline (red), the theoretical result after passive erasure conversion (blue), the upper bound of $Q_p$ in (a) (yellow), the corresponding lower bound of $\sqrt{1/Q_p}$ in (b) (orange), and the SQL (green) are shown. In this setting, all noisy terms fall into $\mathcal{T}_{\mathrm{conv}}$, so the passive erasure-conversion scheme removes all erasure-convertible components and reaches the bound. Across a broad range of $p$, the result after passive erasure conversion substantially outperforms the noisy baseline. In particular, at $p=0.5$, the noisy baseline carries no information, whereas the result after passive erasure conversion still preserves half of the ideal QFI.}
  \label{fig:phase-product}
\end{figure}
Fig.~\ref{fig:phase-product} shows the dependence of $Q_p$ on $p$ at $N=6$, together with the dependence of $\sqrt{1/Q_p}$ on $N$ at $p=0.4$, corresponding to a relatively high phase-damping probability. The result after passive erasure conversion substantially outperforms the noisy baseline over a broad range of $p$. In particular, at $p=0.5$, where the noisy baseline has vanishing QFI, our framework still preserves half of the ideal QFI.

\paragraph{Amplitude damping.}
We next consider the amplitude damping channel, which serves as a representative non-Pauli noise model. The Pauli-basis re-expression of the amplitude damping channel is given in Methods. The noisy baseline is
$Q_\gamma^{\rm noisy}(N,\gamma)=N(1-\gamma)$,
while passive erasure conversion gives
$Q_\gamma^{\rm er}(N,\gamma)=N\,\frac{1-\gamma}{1-\gamma/2}$.
The corresponding upper bound is
$Q_\gamma^{\rm ub}(N,\gamma)=N\Bigl(1-\frac{\gamma}{2}\Bigr)$.
\begin{figure}[H]
  \centering
  \newcommand{\subw}{0.48\linewidth}

  \begin{subfigure}[t]{\subw}
    \centering
    \includegraphics[width=\linewidth]{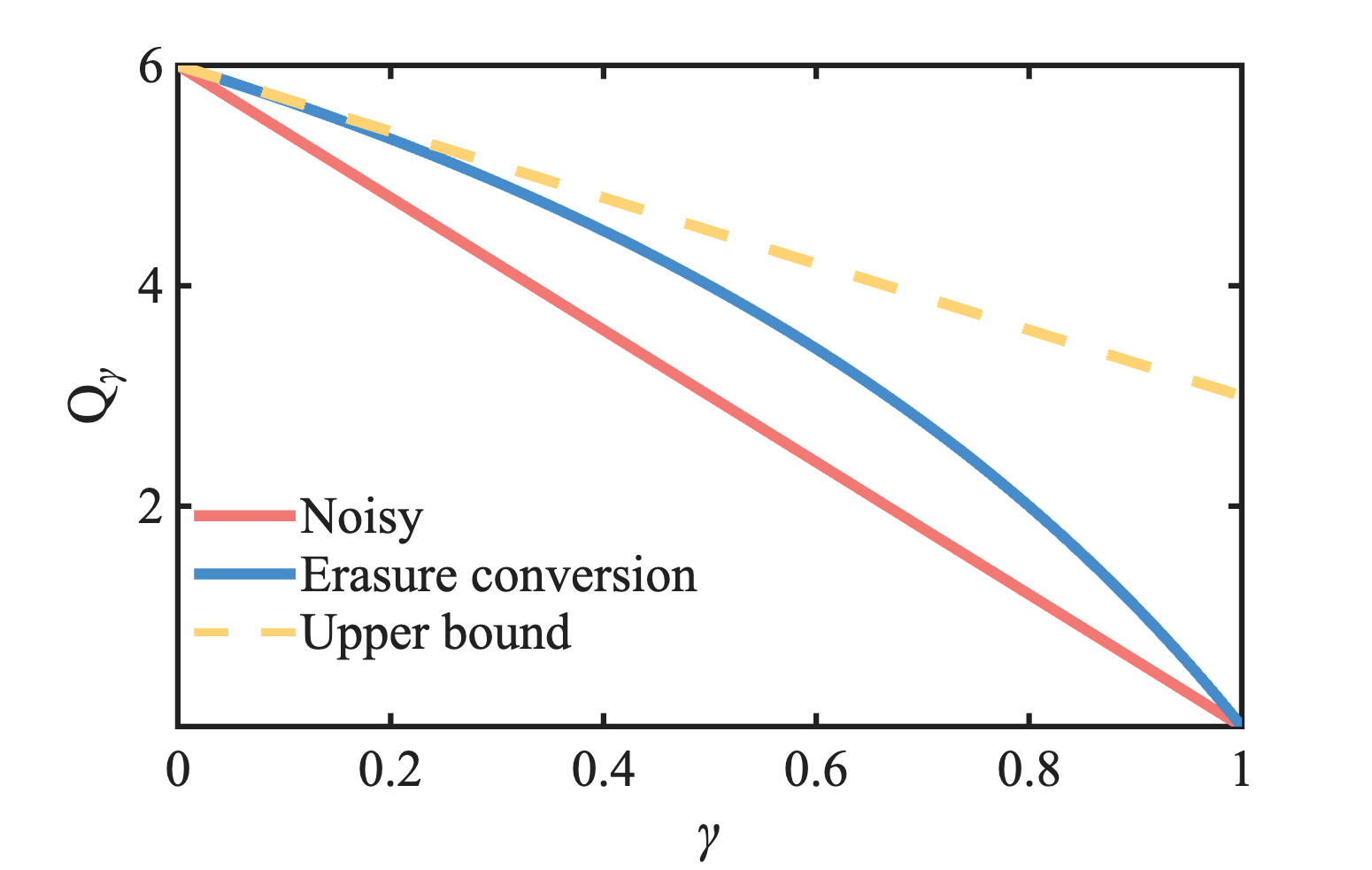}
    \caption{$Q_\gamma$ versus damping strength $\gamma$ at $N=6$.}
  \end{subfigure}\hfill
  \begin{subfigure}[t]{\subw}
    \centering
    \includegraphics[width=\linewidth]{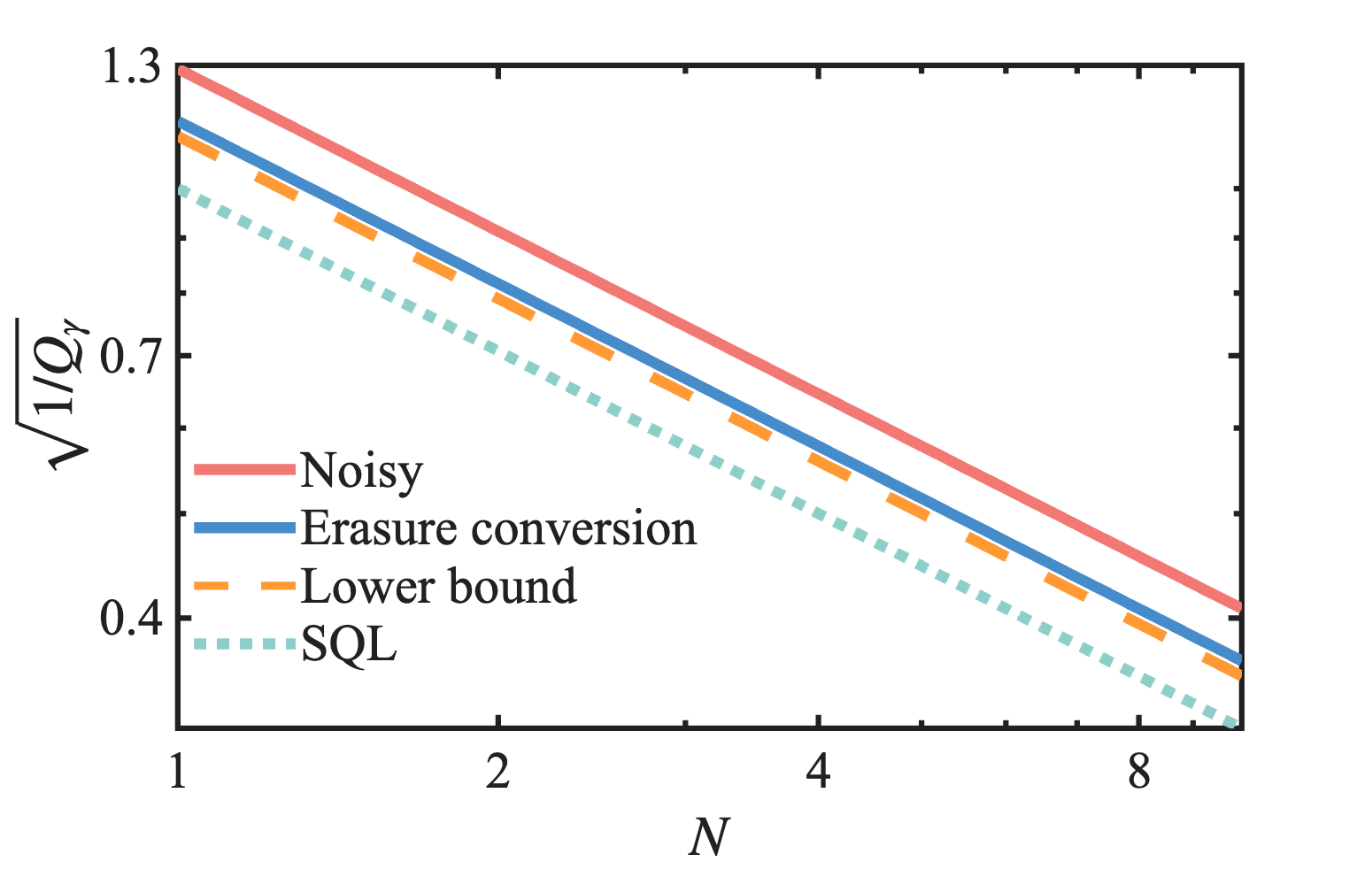}
    \caption{$\sqrt{1/Q_\gamma}$ versus $N$ on log\text{-}log axes at $\gamma=0.4$.}
  \end{subfigure}
  \caption{\textbf{Amplitude damping with residual non-convertible noise components.}
The noisy baseline (red), the theoretical result after passive erasure conversion (blue), the upper bound of $Q_\gamma$ in (a) (yellow), the corresponding lower bound of $\sqrt{1/Q_\gamma}$ in (b) (orange), and the SQL (green) are shown. Because amplitude damping contains residual non-convertible noise components, the curves after passive erasure conversion do not coincide with the bounds. Nevertheless, the result after passive erasure conversion consistently outperforms the noisy baseline and remains much closer to the SQL.}
  \label{fig:amp-product}
\end{figure}

Amplitude damping leaves residual non-convertible noise components in the retained outcome, so $Q_\gamma$ no longer coincides with the upper bound. Nevertheless, the result after passive erasure conversion consistently outperforms the noisy baseline and remains close to the SQL. The dependence of $Q_\gamma$ on $\gamma$ at $N=6$, together with the dependence of $\sqrt{1/Q_\gamma}$ on $N$ at $\gamma=0.4$, is shown in Fig.~\ref{fig:amp-product}.

We also analyze the phase damping case for GHZ probes and the corresponding scaling behavior. These results are presented in the Supplementary Material.

\subsection*{Experimental demonstration}
We experimentally realize passive erasure-conversion on a photonic platform, where polarization serves as the sensing qubit and OAM serves as the ancilla. In this implementation, the signal and the non-convertible part of the noise are restored to the fundamental Gaussian mode, while the converted outputs are routed to higher-order OAM modes for spatial filtering.

At the encoding stage, a Q-plate with topological charge \(l=8\) is used to implement
\begin{equation}
V_1 = \left ( \sigma_{z} \otimes \mathbb{I}_{o} \right ) \left ( |L\rangle\langle L| \otimes e^{i l \hat{\theta}} + |R\rangle\langle R| \otimes e^{-i l \hat{\theta}} \right )
= |R\rangle\langle L| \otimes e^{i l \hat{\theta}} + |L\rangle\langle R| \otimes e^{-i l \hat{\theta}},
\end{equation}
which encodes the initial state \(|H\rangle\otimes|0\rangle\) into the encoded state
$|\psi_{\text{enc}}\rangle = \tfrac{1}{\sqrt{2}}\big(|R\rangle|l\rangle + |L\rangle|-l\rangle\big)$.
For the sensing generator \(G=\sigma_y/2\), whose eigenbasis is \(\{|R\rangle,|L\rangle\}\), a general polarization noise channel
can be decomposed into the non-convertible set
$\mathcal T_{\mathrm{nconv}}
=
\Bigl\{
\sigma_i(\cdot)\sigma_j\;\Big|\; \sigma_i,\sigma_j\in\{I,\sigma_y\}
\Bigr\}$,
and the erasure-convertible set
$\mathcal T_{\mathrm{conv}}
=
\Bigl\{
\sigma_i(\cdot)\sigma_j\;\Big|\; \sigma_i\; \text{or}\ \sigma_j\in\{\sigma_x,\sigma_z\}
\Bigr\}$,
as detailed in Methods. The terms in \(\mathcal T_{\mathrm{nconv}}\) do not induce transitions between the sensing-basis states $|R\rangle $ and $ |L\rangle$, and therefore remain in the same output as the signal, forming the residual noise. By contrast, the terms in \(\mathcal T_{\mathrm{conv}}\) do induce such transitions and can therefore be converted into erasures and removed from the retained Gaussian output.

\begin{figure}[H]
    \centering
    \includegraphics[width=0.5\textwidth]{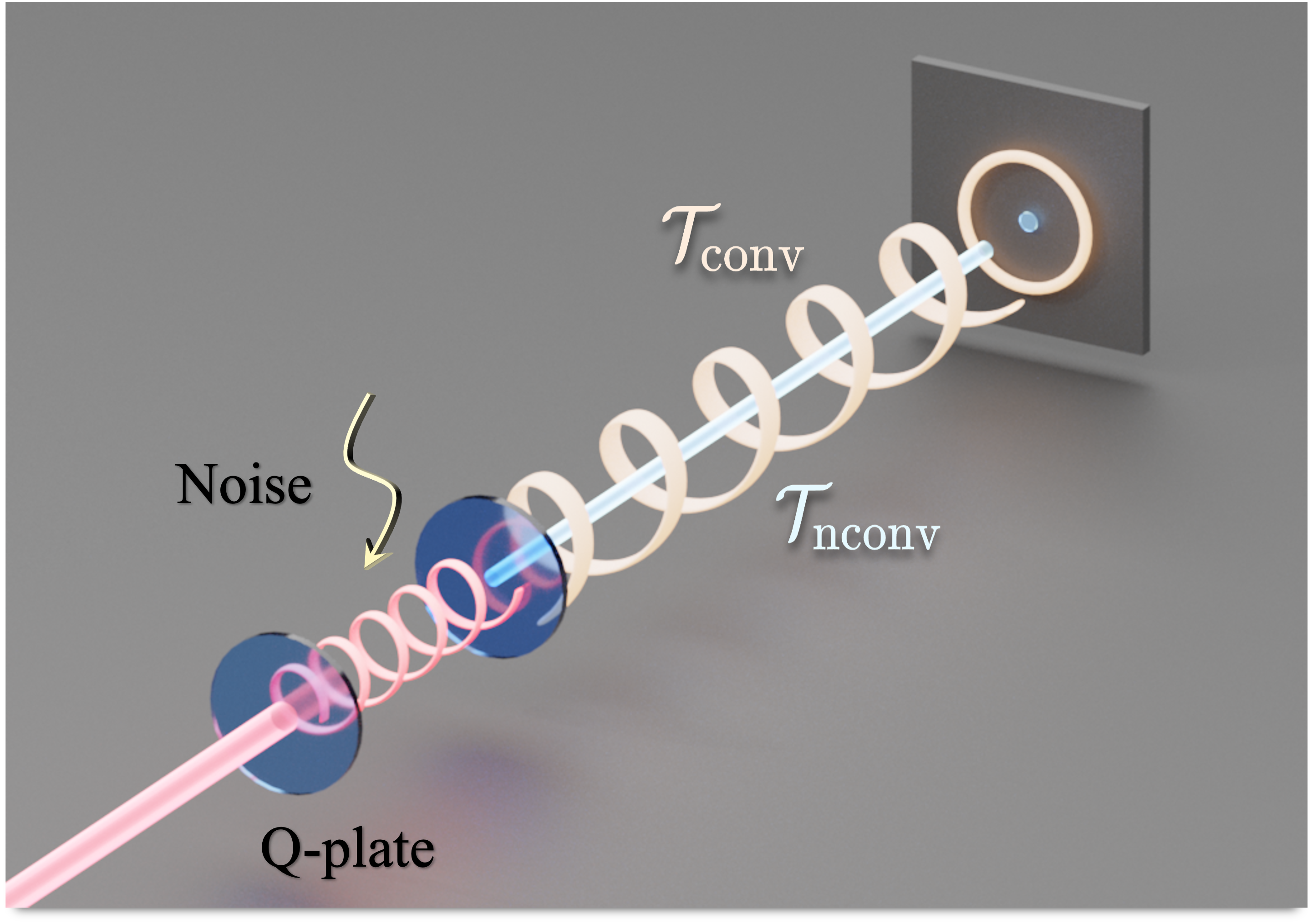} 
\caption{\textbf{Experimental implementation of passive erasure conversion.}
The polarization qubit is coupled to the OAM  by a Q-plate with topological charge \(l=8\), such that the two circular polarizations acquire opposite OAM modes. After the noisy sensing evolution, a second identical Q-plate performs decoding. In this process, the signal together with the non-convertible  noise components is restored to the fundamental Gaussian mode (\(l=0\)), whereas converted outputs are routed to higher-order OAM modes (\(l=\pm16\)). These higher-order modes are then removed by spatial filtering.}
    \label{fig:erasure_setup}
\end{figure}

After the noisy evolution, a second identical Q-plate is applied to implement the decoding operation \(V_2\). Under this decoding, the signal together with \(\mathcal T_{\mathrm{nconv}}\) is restored to the Gaussian mode \((l=0)\), while the converted outputs arising from \(\mathcal T_{\mathrm{conv}}\) are routed to higher-order OAM modes \((l=\pm16)\), as illustrated in Fig.~\ref{fig:erasure_setup}. The experimental demonstration here is presented for noiseless OAM. However, the scheme can also tolerate OAM noise, such as turbulence, provided that it does not bring the erroneous outputs back to the Gaussian mode (see Supplementary).

\begin{figure}[H]
    \centering
        \includegraphics[width=0.78\textwidth]{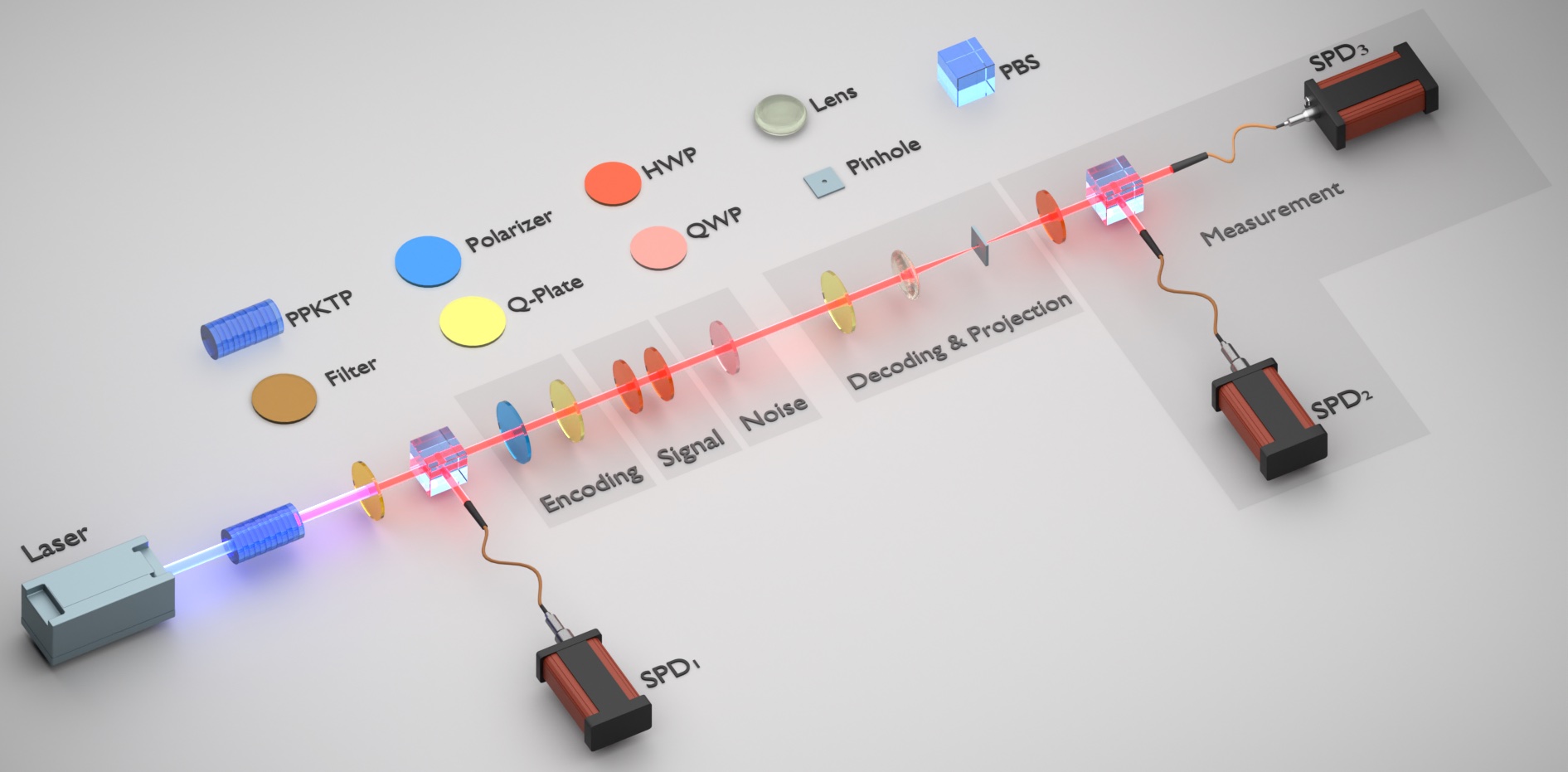}
\caption{
\textbf{Experimental setup.} The passive erasure-conversion consists of four stages: state encoding, noisy signal evolution, decoding and ancilla projection, and final-state measurement. Photon pairs at 810~nm are generated via SPDC in a PPKTP crystal pumped at 405~nm. Idler photons are separated by a PBS to herald the signal photons. The signal photon is initialized in horizontal polarization, with polarization and OAM serving as the sensing qubit and ancilla, respectively. Q-plate~1 encodes the state into OAM modes with \(l=\pm 8\). Two HWPs implement the unitary \(U=e^{-i\alpha \sigma_y/2}\), and a QWP is used to emulate a mixed \(\sigma_x\)\text{-}\(\sigma_z\) noise channel with an erasure-convertible weight of 0.5. Q-plate~2 performs decoding, under which the signal is restored to the Gaussian mode \((l=0)\), whereas the erasure-convertible output states are mapped to higher-order OAM modes \((l=\pm 16)\). A lens and a 100~\(\mu\)m pinhole isolate the Gaussian mode, thereby implementing the ancilla projection and rejecting the higher-order modes. Final measurement is performed with a PBS and a HWP. Coincidences between SPD1\text{-}SPD2 and SPD1\text{-}SPD3 yield the photon statistics under orthogonal projections.
}
    \label{fig:erasure_setup_S}
\end{figure}

The effectiveness of passive erasure-conversion is demonstrated in a proof-of-principle single-photon experiment, as shown in \Cref{fig:erasure_setup_S}. Photon pairs are generated via type-II degenerate spontaneous parametric down-conversion in a periodically poled potassium titanyl phosphate (PPKTP) crystal pumped by a 405\,nm continuous-wave laser. The signal and idler photons at 810\,nm are separated by a polarizing beam splitter 

\noindent (PBS). The signal photon is initialized in the state \(|H\rangle \otimes |0\rangle\), where \(|0\rangle\) denotes the Gaussian mode, and the idler photon is detected by a single-photon detector (SPD1) that acts as a herald. After encoding with Q\mbox{-}plate~1, the phase parameter \(\alpha \in [0,2\pi)\) is imprinted on the polarization qubit by
\(U(\alpha)=e^{-i\alpha \sigma_y/2}\) realized using two cascaded half-wave plates (HWPs). The resulting state is
$|\psi_\alpha\rangle = \frac{1}{\sqrt{2}}\big(|R\rangle|l\rangle + e^{i\alpha}|L\rangle|-l\rangle\big)$.

As discussed above, a general polarization noise channel can be decomposed into \(\mathcal T_{\mathrm{conv}}\) and \(\mathcal T_{\mathrm{nconv}}\). 
To enable a direct comparison between the experimental results and the theoretical predictions, we use a QWP to emulate a Pauli-noise channel containing mixed \(\sigma_x\)\text{-}\(\sigma_z\) noise components in \(\mathcal T_{\mathrm{conv}}\), with erasure-convertible weight 0.5.
Under the ancilla projection, the QWP implementation yields the same measurement statistics and root-mean-square error (RMSE) as the corresponding channel with mixed-state output (see Methods).

Because of diffraction, the Gaussian output and the higher-order OAM outputs become distinguishable only after sufficient propagation (see Supplementary). To emulate far field filtering, we place a lens after the second Q\mbox{-}plate and a 100\,\textmu m pinhole at its focal plane that serves as a spatial filter, passing the \(l=0\) Gaussian mode while rejecting higher-order OAM modes.
This implements the OAM projection \(M_{s} = \mathbb{I}_{p} \otimes |0\rangle_o\langle 0|\).

For parameter estimation, the output state is projected onto the diagonal polarization basis using a HWP set at \(22.5^\circ\) followed by a PBS, which implements the projectors \(|+\rangle_p\langle +|\) and \(|-\rangle_p\langle -|\). Equivalently, this 

\noindent measurement is described by
$\hat{\Pi}_\pm = |\pm\rangle_p\langle\pm| \otimes \mathbb{I}_o$,
and the corresponding probabilities are
\begin{equation}
P_\pm = \langle \psi_f | \hat{\Pi}_\pm | \psi_f \rangle = \tfrac{1}{2}\bigl(1 \mp \sin \alpha\bigr).
\end{equation}
We record coincidence counts between SPD1 and SPD2 and between SPD1 and SPD3 to obtain the statistics under \(\hat{\Pi}_+\) and \(\hat{\Pi}_-\), respectively. These measurements yield an estimate of \(\alpha\), from which we compute the RMSE (see Supplementary).

To evaluate the performance of the passive erasure-conversion, we first conducted experiments with a pinhole inserted for spatial filtering, followed by control measurements without the pinhole. The coincidence window was set to 2\,ns. Photon counts were recorded for five integration times of 0.20, 0.25, 0.50, 0.75, and 1.00\,s. Each setting was repeated 220 times. For statistical analysis, the 220 data sets were grouped into 

\begin{figure}[H]
    \centering
\includegraphics[width=0.66\linewidth]{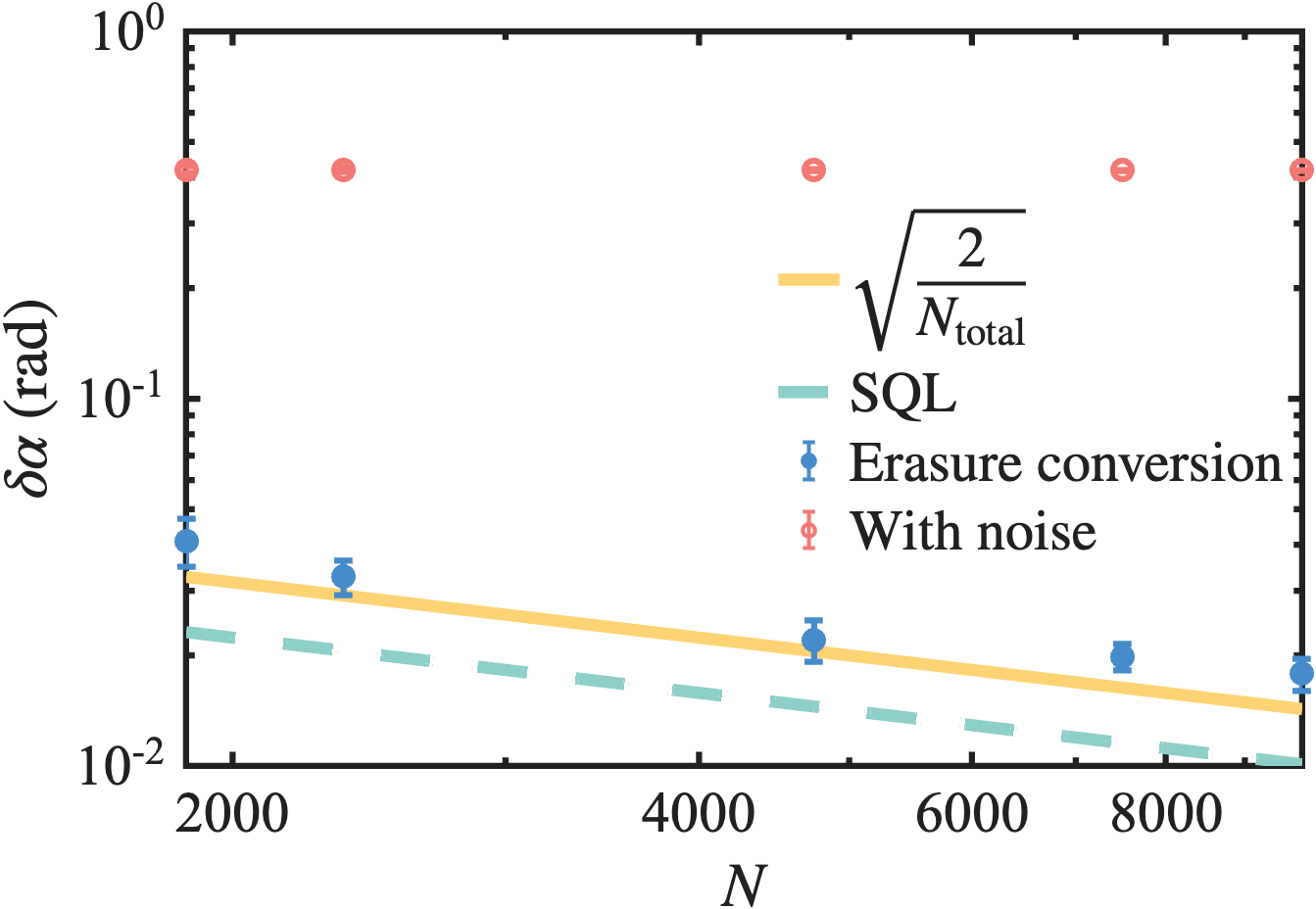}
\caption{\textbf{Experimental results.} 
RMSE of the estimated parameter versus the mean photon number per group. 
The blue circles denote experimental data acquired using the passive erasure-conversion scheme, whereas the red circles represent control measurements without spatial filtering. The yellow solid line indicates the theoretical precision after passive erasure-conversion, while the green dashed line marks the SQL. The passive erasure-conversion data closely track the theoretical curve, confirming the recovery of measurement precision. In contrast, the control data reveal a  persistent deviation caused by the unremoved noise, resulting in consistently elevated RMSE values.
}
    \label{fig:result1}
\end{figure}

\noindent 5 batches of 44 trials each. Experimental results with error bars are shown in \Cref{fig:result1}. Blue circles represent the experimental results with passive erasure-conversion scheme, while red circles correspond to the control condition without spatial filtering. The yellow solid curve shows the theoretical precision after passive erasure-conversion, given by \(\delta\alpha \ge \sqrt{\frac{1}{N_{\mathrm{total}}/2}}\), where \(N_{\mathrm{total}}\) is the total number of photons participating in the parameter encoding stage. The green dashed line indicates the SQL. 
The passive erasure-conversion experimental data closely track the theoretical precision curve, demonstrating that the scheme restores the measurement precision to the expected level. In contrast, the control data exhibit a pronounced deviation due to the unremoved QWP-emulated noise, resulting in consistently high RMSE values.

\section*{Discussion}
Erasure noise is favorable for quantum sensing because erroneous outcomes can be identified and excluded from the data used for parameter estimation. However, realistic sensing noise rarely appears directly as erasures. Instead, it often acts within the same sensing Hilbert space as the signal, making the erroneous contribution hidden in the sensing outcome. This work addresses this gap by establishing a necessary and sufficient erasure-conversion condition for general in-space noise, which determines which noise components can potentially be converted into erasures without damaging the signal. This condition separates a general in-space noise channel into an erasure-convertible sector $\mathcal T_{\rm conv}$ and a non-convertible sector $\mathcal T_{\rm nconv}$ with respect to the sensing generator $G$, rather than the detailed noise model. Guided by this condition, we construct a passive scheme that converts the terms in $\mathcal T_{\rm conv}$ into erasures and removes them from the retained sensing outcome. Our theoretical and experimental results show that the result after passive erasure conversion can substantially restore measurement precision, especially when the weight of the non-convertible noise component is low. Since the execution of passive erasure conversion does not require a detailed noise model, active control, real-time feedback, or precise timing, it provides a lower-overhead route to robust quantum sensing.

Even when the sensing generator can be optimally chosen, the residual non-convertible noise components remain the fundamental obstacle to full precision recovery. The performance of passive erasure conversion is ultimately determined by the weight of the noise-channel components classified into $\mathcal T_{\rm conv}$. In the present framework, the converted erasures are not further used, but are removed by the final erasure readout. This makes the scheme much easier to implement than full quantum error correction, but also means that any information potentially carried by the removed outputs is not used. Whether part of the information contained in the removed outputs can be further extracted or recycled to improve the achievable precision remains an important question for future investigation.

Passive erasure conversion is naturally compatible with other noise-mitigation strategies. It can serve as a lightweight front-end that reshapes uncontrolled physical noise into a structured combination of erasure-convertible and non-convertible parts, and may be combined with more sophisticated protocols such as error-detecting or error-correcting codes, dynamical decoupling, or adaptive estimation strategies. Such hybrid schemes may further exploit information contained in the removed outputs, suppress the remaining non-convertible components, or optimize the sensing generator so that a larger fraction of the noisy terms are classified as erasure-convertible. These possibilities point to broader routes for robust quantum sensing in realistic noisy environments.

\section*{Methods}\label{results}
\subsection*{Operator-basis expansion of a general in-space noise channel}
We consider a $d$-dimensional sensing system with Hilbert space $\mathcal H_S$
and operator space $\mathcal L(\mathcal H_S)$, where $\dim \mathcal H_S = d$. An in-space noise channel is described as a CPTP map
$\mathcal N:\mathcal L(\mathcal H_S)\rightarrow \mathcal L(\mathcal H_S)$.
Let $\{E_\mu\}$ be a complete Hilbert\text{-}Schmidt orthonormal operator basis satisfying $\langle E_\mu, E_\nu\rangle_{\mathrm{HS}} = \delta_{\mu\nu}$,
with 
$\langle A,B\rangle_{\mathrm{HS}} := \mathrm{Tr}(A^\dagger B)$.
Any such CPTP map $\mathcal N$ admits a
Kraus representation $\mathcal N(\cdot) = \sum_r K_r (\cdot) K_r^\dagger$,
and expanding each Kraus operator as $K_r = \sum_\mu a_{r\mu} E_\mu$ yields
the $\chi$-matrix representation
\begin{equation}
\mathcal N(\cdot) = \sum_{\mu,\nu} \chi_{\mu\nu}\, E_\mu (\cdot) E_\nu^\dagger,
\label{eq:chi_expansion_methods}
\end{equation}
with $\chi_{\mu\nu} := \sum_r a_{r\mu} a_{r\nu}^*$.
The matrix $\chi$ is positive semidefinite, and the trace-preserving condition
$\sum_r K_r^\dagger K_r = I$ becomes
\begin{equation}
\sum_{\mu,\nu} \chi_{\mu\nu}\, E_\nu^\dagger E_\mu = I.
\label{eq:TP_condition_methods}
\end{equation}
Eqs~\eqref{eq:chi_expansion_methods} and \eqref{eq:TP_condition_methods}
characterize a CPTP channel in the orthonormal operator basis $\{E_\mu\}$~\cite{Choi1975,Jamiolkowski1972,NielsenChuang2010}.

\subsection*{Optional optimization of the sensing generator \(G\)}

The performance of passive erasure-conversion improves with the weight of erasure-convertible terms. When equivalent sensing bases are available, the induced erasure-convertible set \(\mathcal T_\mathrm{conv}\) is not unique and can be optimized through an appropriate choice of the equivalent sensing generator. Formally, let \(R\) denote an allowed unitary transformation on \(\mathcal H_S\), and define the corresponding equivalent sensing generator as \(G'(R)=RGR^\dagger\). 
For each such \(R\), the noise channel is re-expressed with respect to \(G'(R)\) as
\(\mathcal N(\cdot)=\sum_{\mu,\nu}\chi_{\mu\nu}^{(R)}\,E_{\mu}^{(R)}(\cdot)E_{\nu}^{(R)\dagger}\),
which induces a corresponding separation of the channel terms into
\(\mathcal T_\mathrm{conv}^{(R)}\) and \(\mathcal T_\mathrm{nconv}^{(R)}\).
We define the total coefficient weight of the erasure-convertible channel terms as
$W_{\rm conv}(R)
=\sum_{\mu,\nu}^{\prime}\left|\chi_{\mu\nu}^{(R)}\right|^2$,
where the prime indicates that the summation is restricted to index pairs whose corresponding channel terms $E_{\mu}^{(R)}(\cdot)E_{\nu}^{(R)\dagger}$ belong to $\mathcal T_\mathrm{conv}^{(R)}$.
One may then choose
$R^\star\in\arg\max_R W_{\rm conv}(R)$.
This choice maximizes the total coefficient weight of the erasure-convertible channel terms under the allowed reorientation of the sensing generator.
This reorientation increases the portion of noise converted into erasures while leaving the noiseless QFI unchanged. Importantly, this optimization is optional and is not required for the execution of the passive erasure-conversion scheme.

\subsection*{Pauli-basis re-expression of amplitude damping}
The Kraus operators of the amplitude damping channel are
\begin{equation}
K_0=
\begin{pmatrix}
1&0\\
0&\sqrt{1-\gamma}
\end{pmatrix},
\quad
K_1=
\begin{pmatrix}
0&\sqrt{\gamma}\\
0&0
\end{pmatrix}.
\end{equation}
For the sensing generator \(G=\sigma_z/2\), the Kraus operators of amplitude damping can be re-expressed in the Pauli basis as
\begin{equation}
K_0=
\frac{1+\sqrt{1-\gamma}}{2}\,I+
\frac{1-\sqrt{1-\gamma}}{2}\,\sigma_z,
\quad
K_1=\sqrt{\gamma}\,\frac{\sigma_x+i\sigma_y}{2}.
\end{equation}
Accordingly, the amplitude damping channel can be expanded as
\begin{equation}
\mathcal N_\gamma(\rho)
=
a^2\rho
+ab(\sigma_z\rho+\rho \sigma_z)
+b^2 \sigma_z\rho \sigma_z
+\frac{\gamma}{4}\bigl(\sigma_x\rho \sigma_x+\sigma_y\rho \sigma_y+i\sigma_y\rho \sigma_x-i\sigma_x\rho \sigma_y\bigr),
\end{equation}
where
$a=\frac{1+\sqrt{1-\gamma}}{2}, b=\frac{1-\sqrt{1-\gamma}}{2}$.

For \(G=\sigma_z/2\), one has \(\mathcal D_G=\mathrm{span}\{I,\sigma_z\}\) and \(\mathcal O_G=\mathrm{span}\{\sigma_x,\sigma_y\}\). Therefore, the terms \(\rho\), \(\sigma_z\rho\), \(\rho \sigma_z\), and \(\sigma_z\rho \sigma_z\) belong to \(\mathcal T_{\mathrm{nconv}}\), while \(\sigma_x\rho \sigma_x\), \(\sigma_y\rho \sigma_y\), \(\sigma_y\rho \sigma_x\), and \(\sigma_x\rho \sigma_y\) belong to \(\mathcal T_{\mathrm{conv}}\).

\subsection*{Channel-term analysis of polarization noise}
In the photonic experiment, polarization serves as the sensing qubit, while OAM serves as the ancilla. We choose the sensing generator as
$G=\sigma_y/2$, whose eigenbasis is \(\{|R\rangle,|L\rangle\}\). Relative to this sensing basis, the operator space is decomposed into the diagonal subspace $\mathcal D_G=\mathrm{span}\{|R\rangle\langle R|,\ |L\rangle\langle L|\}$ and the off-diagonal subspace $\mathcal O_G=\mathrm{span}\{|R\rangle\langle L|,\ |L\rangle\langle R|\}$.
Accordingly, the Pauli operators satisfy $I,\ \sigma_y\in\mathcal D_G$ and $\sigma_x,\ \sigma_z\in\mathcal O_G$.

A general polarization noise channel can therefore be written as $\mathcal N(\rho)=\sum_{i,j\in\{0,x,y,z\}} \chi_{ij}\,\sigma_i\rho\sigma_j$,
where \(\sigma_0\equiv I\), and the channel terms are classified into the non-convertible set $\mathcal T_{\mathrm{nconv}}
=
\bigl\{
\sigma_i\rho\sigma_j\ \big|\ \sigma_i,\sigma_j\in\{I,\sigma_y\}
\bigr\}$ and the erasure-convertible set $
\mathcal T_{\mathrm{conv}}
=\bigl\{\sigma_i\rho\sigma_j\ \big| \sigma_i\; \text{or}\; \sigma_j\in\{\sigma_x,\sigma_z\}\bigr\}$.

The noisy terms in \(\mathcal T_{\mathrm{nconv}}\) do not induce transitions between the sensing-basis states $|R\rangle$ and $|L\rangle$, and therefore have the same structure as the signal. Consequently, they remain in the retained output together with the signal and constitute the residual noise that cannot be removed without also removing the signal. In the experiment, this retained output is mapped back to the Gaussian mode.

Within \(\mathcal T_{\mathrm{conv}}\), the terms can be divided into two categories according to their distinct physical manifestations in the experiment: same-operator terms, of the form \(\sigma_i\rho\sigma_i\), which correspond to actual output states that are mapped to higher-order OAM modes; and cross terms, of the form \(\sigma_i\rho\sigma_j\,(i\neq j)\), which describe coherence between different output states. Once the higher-order OAM outputs are removed by spatial filtering, the associated cross terms vanish accordingly.

\subsection*{QWP emulation of a channel with fixed erasure-convertible weight}

To stably emulate a mixed \(\sigma_x\)\text{-}\(\sigma_z\) Pauli-noise channel, we use a QWP whose fast-axis orientation is arbitrary in the experiment.
After the final ancilla projection, the QWP implementation is equivalent, at the level of retained measurement statistics, to the corresponding channel that produces a mixed-state output with an erasure-convertible weight of 0.5.

The phase retardation of the QWP is $\delta=\pi/2$. When the fast axis forms an angle $\beta$ with the horizontal polarization, the QWP implements the polarization unitary
\begin{equation}
U_{\mathrm{QWP}}(\beta)=\frac{1}{\sqrt{2}}\left[\mathbb{I}-i(\sin2\beta\,\sigma_x+\cos2\beta\,\sigma_z)\right]
=\frac{1}{\sqrt{2}}\left(\mathbb{I}-i\sigma_n\right),
\label{eq:Uqwp}
\end{equation}
where $\sigma_n=\sin2\beta\,\sigma_x+\cos2\beta\,\sigma_z$.

Applying the QWP to the probe gives
\begin{equation}
\rho_f=U_{\mathrm{QWP}}(\beta)\rho U_{\mathrm{QWP}}^{\dagger}(\beta)
=\tfrac{1}{2}\rho+\tfrac{i}{2}\rho\sigma_n-\tfrac{i}{2}\sigma_n\rho+\tfrac{1}{2}\sigma_n\rho\sigma_n. 
\label{eq:rho_f1}
\end{equation}
Eq.~(\ref{eq:rho_f1}) shows that the QWP-induced channel contains equal-weight signal-preserving and erasure-convertible components, with the signal-preserving component \(I\rho I\) carrying weight \(1/2\) and the erasure-convertible components carrying the remaining weight \(1/2\).
The $\beta$-dependence is carried by $\sigma_n$, while the total erasure-convertible weight remains fixed at $1/2$.

Under the same projection, the QWP-induced output is observationally equivalent to the channel $\mathcal{F}_{1/2}$, which produces a mixed-state output with an erasure-convertible-component weight of $0.5$, namely $\mathcal{F}_{1/2}(\rho)=\tfrac{1}{2}\rho+\tfrac{1}{2}\sigma_n\rho\sigma_n$. 
Because the final ancilla projection removes the terms in \(\mathcal T_{\mathrm{conv}}\) from the retained outcome, both channels yield the same retained result, \(\tfrac{1}{2}\rho\), as well as the same measurable statistics, including count rates and the RMSE scaling with sample size.

\section*{Data availability}
The data that support the findings of this study are available within the paper and its Supplementary Information. Any additional information is available from the corresponding authors upon request.

\section*{Acknowledgements}
This work was supported by the National Natural Science Foundation of China (No. 62471289), Natural Science Foundation of Shanghai (24ZR1432900), Quantum Science and Technology-National Science and Technology Major Project (No.2021ZD0300703) and Shanghai Municipal Science and Technology Major Project (Grant No. 2019SHZDZX01).
\section*{Author Contributions}
J.H. and G.Z. supervised the project. X.L. and J. H. constructed the theoretical model and carried out the experiments with assistance from Z.C., B.X., Y.L. and H.L., X.L. analyzed the data. X.L. and J.H. wrote the manuscript. All authors have read and approved the final version of the manuscript.

\section*{Declaration of Interests}
The authors declare no competing interests.

\nocite{Liu2020QFIM,Rezakhani2019ContinuityQFI,Farias2015,Goodman2005FourierOptics}
\bibliography{bibliography}
\bibliographystyle{naturemag}

\end{document}